%% file: main.tex
\documentclass[sigconf, 10pt, nonacm]{acmart}

\setcopyright{none}

\makeatletter
\let\@affiliationfont\@authorfont
\makeatother

\AtBeginDocument{}

\usepackage{amsmath,amsthm,mathtools}
\usepackage{graphicx}
\graphicspath{{fig/}{tex/fig/}}
\usepackage{booktabs}
\usepackage{array}
\usepackage{subcaption}
\usepackage[shortlabels]{enumitem}
\usepackage{tikz}
\usetikzlibrary{arrows.meta,calc,positioning,fit,backgrounds}

\begin{document}

\title{PACE: A Playback-Aligned Context Engine for LLM-Based Full-Duplex Voice Dialogue}

\author{Shibo Wang\textsuperscript{*}, Zicheng Zhang\textsuperscript{*}, Libo Wang\textsuperscript{\textdagger}, and Junfeng Ma}
\affiliation{%
  \institution{Alibaba Group}
  \country{}}

\renewcommand{\shortauthors}{Shibo Wang, Zicheng Zhang et al.}

\begin{abstract}
LLM-based full-duplex voice services allow users to speak while the assistant
is responding. Because servers can generate output and advance dialogue state
faster than clients can play it, subsequent user speech may be interpreted
based on content the user never heard. We call this failure
\emph{Generative Context Mis-anchoring} (GCM).
To address GCM issues, we present PACE, a provider-independent middleware layer that
anchors model-facing context to the client playback boundary, a
system-observable proxy for what the user could have heard. After an
interruption, PACE repairs this context to exclude assistant content that
never reached playback, while preserving low-latency generation across
heterogeneous voice runtimes.

We implement PACE's audio-only projection path end to end in a browser-based
realtime voice assistant using a black-box speech model, without modifying the
model service.
We also construct GCM-Bench\footnotemark, a new controlled benchmark dataset
of 108 playback-relative referent-anchoring cases. On GCM-Bench, PACE raises
Referent Anchoring Accuracy from 25.0\% to 96.3\% over a cancellation-only
baseline. On 200 Full-Duplex-Bench v1 interruption samples, it preserves
interruption response quality.
These results show that grounding model-facing context in actual playback is a
practical way to maintain consistency in full-duplex voice dialogue.
\end{abstract}

% ...


\maketitle
\begingroup
\renewcommand{\thefootnote}{*}
\footnotetext{Co-first authors.\quad\textsuperscript{\textdagger}Corresponding author.}
\endgroup
\footnotetext[1]{Dataset: \url{https://github.com/CodeForZzc/GCM-Bench}.}

\input{tex/introduction}
\input{tex/background}
\input{tex/algo_design}
\input{tex/implementation}
\input{tex/dataset}
\input{tex/evaluation}
\input{tex/related_work}
\input{tex/discussion}
\input{tex/conclusion}

\bibliographystyle{ACM-Reference-Format}
\bibliography{reference}

\end{document}

%% file: tex/introduction.tex
\section{Introduction}\label{sec:intro}

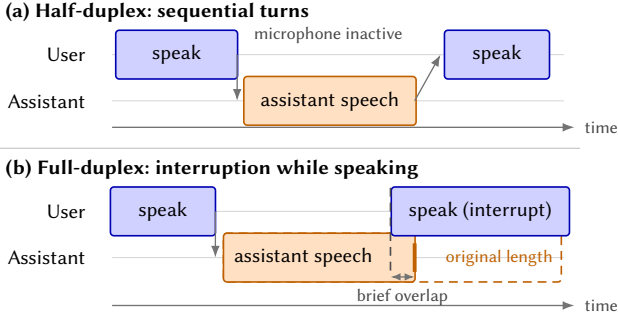
\begin{figure}[t]
  \centering
  \resizebox{\columnwidth}{!}{\input{fig/duplex-comparison}}
  \caption{Half- and full-duplex voice interaction timelines. In full-duplex
  dialogue, user speech can overlap with and interrupt assistant output.}
  \Description{Two aligned timelines compare user and assistant speech. The
  half-duplex timeline contains non-overlapping alternating turns, whereas the
  full-duplex timeline shows a user interruption, a short overlap, the point at
  which assistant output stops, and a dashed continuation representing the
  remaining speech that would have played without interruption.}
  \label{fig:duplex-comparison}
  \vspace{-0.2cm}
\end{figure}

LLM-based voice assistants are moving beyond strict turn-taking toward
full-duplex dialogue, in which the service continues listening while assistant
audio is generated and played. By allowing user speech to overlap with
assistant output, this interaction pattern supports interruptions,
backchannels, and clarifications without requiring users to wait for a response
to finish
\cite{minicpm2026,moshi2024,geminiLiveDocs}, as illustrated in
Fig.~\ref{fig:duplex-comparison}.

Research on full-duplex dialogue has largely focused on coordinating this
overlap. Learned full-duplex models represent overlapping speech,
interjections, and interruptions without strict speaker-turn segmentation
\cite{syncllm2024,moshi2024,lslm2024,minicpm2026,ntpp2025}.
Modular systems use explicit dialogue states, personalized VAD, semantic
end-of-turn detection, and dialogue managers to control cascaded or
semi-cascaded ASR--LLM--TTS pipelines
\cite{duplexconversation2022,fullduplexscheme2024,flexduo2025,llmdm2025,duplexcascade2026,fireredchat2025,unitagent2026}.
Benchmarks and production APIs further emphasize pause handling,
backchanneling, endpointing, and interruption control
\cite{fullduplexbench2025,fdbench2025,fullduplexbenchv22025,mtrduplexbench2025,openaiRealtimeVad,livekitTurns}.
Together, these efforts advance the modeling, control, and evaluation of
overlapping speech and turn-taking.

Yet these advances overlook a basic asymmetry: assistant speech is generated
far faster than it is played back to the user. Because server-side dialogue
state advances with generation, the model may have produced and recorded
several sentences while the client is still playing the first. What the model
assumes the user has received thus diverges from what has actually been
played: a user utterance referring to the sentence just played may be
interpreted against sentences that have not yet been played. Network RTT,
inference queues, packet jitter, browser audio buffers, and local playback
holds further widen this generation--playback gap.

We call this failure mode \emph{Generative Context Mis-anchoring} (GCM). Unlike
an ASR error, a TTS error, or a conventional hallucination, GCM is a distributed
consistency failure: the recognized input and generated output may both be
locally correct, but the input is anchored to the wrong point in the
conversation. For example, a user may interrupt a list of tourist attractions
after hearing the second item and ask, ``How do I get there?'' If the model has
already generated and recorded a fourth item, it may resolve ``there'' to that
unheard attraction and produce a coherent answer to the wrong referent.

Stopping playback and canceling generation prevent additional output from
reaching the user, but do not repair model-facing context that has
already advanced. OpenAI's Realtime API goes further: item truncation removes
unplayed assistant audio and corresponding transcript content from server-side
context~\cite{openaiRealtime}. However, this mechanism is tied to one provider's
conversation-item abstraction and still requires a mapping from client playback
to dialogue state. We discuss these limitations in
Section~\ref{sec:state-surgery}.

To address GCM more generally, we present PACE, the \emph{Playback-Aligned
Context Engine}, a provider-independent middleware abstraction that makes the
mapping from client playback to model-facing dialogue state explicit.
PACE separates two concerns: tracking how far playback has progressed, and
repairing model-facing context through whatever mechanism the runtime offers.
For each assistant turn, a server-side output ledger records the generated
audio, and client playback acknowledgments report the latest sample rendered
for that turn. This position is the physical \emph{playback boundary}: content
before it has been played on the client, whereas content beyond it is
speculative and can still be revoked.

Throughout this paper, we use ``heard''
operationally to denote content that had reached client playback, not to claim
user attention or comprehension. The two sides of the boundary carry
asymmetric evidence: content beyond it could not have been heard, whereas
content before it was rendered and therefore available to be heard. The
playback boundary itself is only a sample position. When the runtime also
provides a reliable mapping from audio time to content, PACE can go one step
further and treat the last span that finished playing before this position as
semantically committed, yielding a conservative semantic commit boundary.
Without such a mapping, PACE claims only which samples were played, not which
content they carried.

On a confirmed interruption, PACE snapshots the playback boundary, closes a
forward gate for new user audio, cancels generation, suppresses late output, and
revokes buffered client audio beyond the boundary. It then projects the played
context before releasing the held user audio. Depending on the runtime, this
projection can rewrite text history in a cascaded system, invoke provider
truncation and cancellation operations, or re-inject recently played audio into
a black-box voice model. The boundary abstraction is therefore decoupled from
the available projection mechanism; exact internal-state restoration is not
assumed for a black-box runtime.

We implement and evaluate PACE's physical, audio-only projection path in a
browser-based realtime voice assistant connected to a black-box speech model,
without modifying the model service. Our new benchmark, GCM-Bench, supplies a
reproducible user-input trajectory for each case, while every trial traverses
the live client--middleware--model path and obtains its playback boundary from
actual browser PlaybackAck events. On GCM-Bench, PACE raises Referent Anchoring
Accuracy from 25.0\% under a cancellation-only baseline to 96.3\%. Separately,
we reuse 200 official Full-Duplex-Bench (FDB) v1 interruption samples for
compatibility and latency evaluation. Both PACE and the baseline respond to
every FDB interruption; the mean response score changes from 4.975 to 4.995.
Requests made while the assistant is not playing follow the original
processing path and incur no additional PACE latency.

This paper makes five contributions.
\begin{enumerate}[leftmargin=*]
  \item We formulate GCM as a distributed consistency failure caused by
  divergence among generated assistant output, client playback, and the
  model-facing dialogue state used for subsequent responses or actions.
  \item We design PACE, a provider-independent middleware abstraction that
  combines a turn-local output ledger, playback acknowledgments, output
  revocation, ordered context projection, optional semantic commitment, and
  gates for consequential actions.
  \item We implement and evaluate the physical, audio-only projection path end
  to end in a browser-based voice assistant connected to a black-box realtime
  model without modifying the model service.
  \item We construct GCM-Bench, a new 108-case benchmark dataset for
  playback-relative referent anchoring, and define its task, metric, and
  evaluation protocol. We separately assess compatibility and latency on 200
  official Full-Duplex-Bench v1 samples.
  \item We show that PACE raises Referent Anchoring Accuracy from 25.0\% to
  96.3\% on GCM-Bench, while preserving interruption response quality on
  Full-Duplex-Bench v1.
\end{enumerate}

%% file: fig/duplex-comparison.tex
\begin{tikzpicture}[
  x=0.96cm,
  y=0.50cm,
  font=\sffamily\scriptsize,
  axis/.style={draw=black!55, line width=0.45pt, -{Latex[length=1.4mm]}},
  guide/.style={draw=black!16, line width=0.35pt},
  user/.style={draw=blue!70!black, fill=blue!18, rounded corners=1.2pt,
               line width=0.55pt},
  assistant/.style={draw=orange!75!black, fill=orange!24, rounded corners=1.2pt,
                    line width=0.55pt},
  overlap/.style={draw=black!55, densely dashed, rounded corners=1.2pt,
                  line width=0.45pt},
  note/.style={font=\sffamily\tiny, text=black!72}
]
  \node[anchor=west, font=\sffamily\bfseries\scriptsize] at (0,5.45)
    {(a) Half-duplex: sequential turns};
  \node[anchor=east] at (1.25,4.45) {User};
  \node[anchor=east] at (1.25,3.35) {Assistant};
  \draw[guide] (1.45,4.45) -- (7.05,4.45);
  \draw[guide] (1.45,3.35) -- (7.05,3.35);
  \draw[axis] (1.45,2.72) -- (7.18,2.72) node[right, note] {time};

  \node[user, minimum width=1.45cm, minimum height=0.58cm] at (2.25,4.45)
    {speak};
  \node[assistant, minimum width=2.05cm, minimum height=0.58cm] at (4.15,3.35)
    {assistant speech};
  \node[user, minimum width=1.25cm, minimum height=0.58cm] at (6.22,4.45)
    {speak};
  \draw[axis] (3.00,4.45) -- (3.00,3.35);
  \draw[axis] (5.20,3.35) -- (5.52,4.45);
  \node[note, anchor=south] at (4.13,4.48) {microphone inactive};

  \draw[black!20, line width=0.45pt] (0,2.22) -- (7.25,2.22);

  \node[anchor=west, font=\sffamily\bfseries\scriptsize] at (0,1.72)
    {(b) Full-duplex: interruption while speaking};
  \node[anchor=east] at (1.25,0.72) {User};
  \node[anchor=east] at (1.25,-0.38) {Assistant};
  \draw[guide] (1.45,0.72) -- (7.05,0.72);
  \draw[guide] (1.45,-0.38) -- (7.05,-0.38);
  \draw[axis] (1.45,-1.52) -- (7.18,-1.52) node[right, note] {time};

  \node[user, minimum width=1.25cm, minimum height=0.58cm] at (2.08,0.72)
    {speak};
  \node[assistant, fill=none, densely dashed, minimum width=4.02cm,
        minimum height=0.58cm, anchor=west] at (2.82,-0.38) {};
  \node[assistant, minimum width=2.28cm, minimum height=0.58cm, anchor=west]
    (reply) at (2.82,-0.38) {};
  \node at (3.82,-0.38) {assistant speech};
  \node[note, text=orange!75!black] at (6.25,-0.38) {original length};
  \node[user, minimum width=2.12cm, minimum height=0.58cm, anchor=west]
    (interrupt) at (4.90,0.72) {speak (interrupt)};
  \draw[axis] (2.73,0.72) -- (2.73,-0.38);

  \draw[black!70, densely dashed, line width=0.55pt]
    (4.90,1.28) -- (4.90,-0.71);
  \draw[orange!75!black, line width=1.3pt]
    (5.20,-0.72) -- (5.20,-0.04);
  \draw[draw=black!55, line width=0.45pt,
        {Latex[length=1.2mm]}-{Latex[length=1.2mm]}]
    (4.90,-0.84) -- (5.20,-0.84);
  \node[note, anchor=north] at (5.05,-0.88) {brief overlap};
\end{tikzpicture}

%% file: tex/background.tex
\section{Background and Problem}\label{sec:problem}

\subsection{Full-Duplex Voice Dialogue}\label{sec:full-duplex}

An LLM-based voice assistant can be organized around two
different interaction paradigms. In a half-duplex or push-to-talk interface, the
application alternates between listening and speaking. The user finishes an
utterance, the system processes it, the assistant speaks, and the microphone is
either closed or treated as inactive until the next user turn. This design gives
the application a simple state machine: each assistant response can be considered
complete once generated, and the next user input is assumed to arrive after that
response has been delivered.

Full-duplex voice dialogue changes this paradigm. The microphone remains active
while assistant audio is being generated, streamed, buffered, and played, so the
user can interrupt, backchannel, hesitate, clarify, or continue speaking during
assistant output~\cite{minicpm2026,moshi2024,geminiLiveDocs}. The application is
therefore no longer a sequence of isolated user and assistant turns. It is a
continuous bidirectional stream in which input capture, model generation,
network delivery, audio playback, and user perception proceed concurrently.

This shift makes full-duplex systems feel more natural, but it also changes what
the application must track. The system must decide not only whether new user
speech is a true interruption, a short acknowledgment, acoustic echo, or
background noise, but also which assistant content had reached playback and
could actually have been heard when that speech occurred. Client playback does
not reveal whether the user attended to or understood that content, but it is
the latest system-observable boundary on what the user could have heard. In
other words, full-duplex dialogue separates \emph{generated} assistant context
from \emph{playback-grounded} assistant context.

\subsection{Generative Context Mis-anchoring}
\label{sec:gcm}

GCM occurs when a model interprets a user
input, chooses a continuation, or commits an external action using assistant
content that has been generated but not yet grounded at the user's playback
boundary. Unlike ASR, TTS, or factual errors, the recognized input and generated
output may both be locally correct; the failure is that the model binds the
input to the wrong point in the conversation. GCM becomes likely when assistant
generation outpaces playback, user speech overlaps buffered or playing audio,
and the next model context includes content beyond what the user has heard.
These conditions can arise in both native speech-to-speech runtimes and cascaded
voice agents, and are amplified by faster-than-realtime TTS, large client buffers,
temporary network spikes, or aggressive server-side streaming.

This mismatch appears in several recurring forms.

\noindent\textbf{Temporal reference
shift.}\quad The user responds to an audible sentence, but the model anchors
the response to a later generated one. For example, if the assistant says
``today is sunny'' and then generates ``tomorrow will rain,'' a user who
interrupts after the first sentence and asks, ``Should I bring an umbrella?''
may receive an answer about tomorrow.

\noindent\textbf{Reference resolution collapse.}\quad Phrases such as ``repeat the
last sentence'' resolve against the last generated sentence rather than the
last audible one. This mismatch is especially damaging for instruction-following
tasks, where the user and the assistant can end up several steps apart.

\noindent\textbf{Confirmation binding error.}\quad A confirmation uttered against an audible
proposal may bind to a later generated revision.
A user may hear ``I can book 10am,'' say ``OK,'' and cause the assistant to
execute an unheard change to 2pm. This form is high-risk because it can affect
tool calls, transactions, and irreversible external actions.

\noindent\textbf{Repair failure.}\quad Even when the user notices the drift, the assistant
cannot explain or correct it, because it does not know which content was unheard. A user may say ``wait,
I was talking about the previous one'' or ``I had not heard that yet,'' but the
assistant may treat the confusion as a new semantic request and compound the
misalignment.

\subsection{Why Transport Feedback Alone Is Insufficient}
\label{sec:transport-alone}

Traditional realtime media systems already expose timing and receiver feedback.
RTP and RTCP, for example, provide timestamps and reception reports for media
delivery~\cite{rfc3550}. Browser audio APIs expose clocks and output latency
information that can help estimate audible playout~\cite{webaudio}. Such
signals are useful for media synchronization: they can indicate whether audio
arrived, how much latency accumulated, and which portion of the waveform has
likely reached the speaker.

For a full-duplex voice LLM service, however, the relevant question is not only
where playback is in the audio stream, but where the user is in the dialogue.
Knowing that playback has reached a particular sample offset does not by itself
reveal which sentence the user heard, which option was offered, or which tool
precondition became audible. The sample offset is sufficient for audio-domain
revocation or re-injection, but semantic commitment requires an additional
mapping from speech playout to dialogue state. Transport-level feedback alone
cannot provide that mapping, because it describes the media stream rather than
the state of the conversation. PACE therefore uses playback as an asymmetric
proxy for hearing: content beyond the boundary could not have been heard, while
content before it had at least been rendered and made available to the user.

\subsection{Limitations of Item Truncation and State Rewinding}
\label{sec:state-surgery}

A natural response to GCM is to delete or rewind the unheard suffix of the
assistant response. Item truncation, as exposed by commercial realtime APIs
(Section~\ref{sec:related}), removes the unheard tail of an assistant turn from
the server-side context~\cite{openaiRealtime}. In a self-hosted transformer
runtime, the analogous lower-level operation is to truncate or restore the
key--value (KV) cache---and, for speech-to-speech models, the associated
acoustic and streaming decoder states---back to the token position that
corresponds to the audible boundary. These mechanisms are useful, but they are
not sufficient as a general grounding method.

\noindent\textbf{Runtime projection.}\quad Item truncation repairs a high-level conversation
object, but the public API does not specify how that repair is projected into the
model runtime. If truncation is realized mainly as a conversation-log rewrite,
the next request may need to re-prefill the modified prefix, weakening prefix
caching. If it is realized by reconstructing or rewinding internal state, the
runtime must expose and maintain finer-grained cache state. In both cases, the
operation still depends on a separate playout-to-dialogue mapping if it is to
restore a textual or semantic state: the system must know which text span,
semantic proposal, or tool precondition corresponds to the audio that the user
could have heard at the playback boundary.

\noindent\textbf{Access and precision.}\quad Hosted realtime voice APIs expose high-level
conversation and response events, not direct handles to KV cache or streaming
decoder state~\cite{openaiRealtime}. Self-hosted systems provide
more control, but the repair boundary must still be placed precisely. A playback
boundary can fall inside a word, token, or audio span, while speech-to-speech
runtimes may carry acoustic state, vocoder state, and decoder state beyond the
text tokens alone. Restoring a coherent dialogue state is therefore more
involved than removing a suffix from a transcript or truncating a single tensor.

\noindent\textbf{Serving cost.}\quad Modern runtimes lean on prefix and prompt caching, paged
attention, and continuous batching to sustain throughput. Mutating cache state
mid-stream to track playback works against these optimizations. Checkpointing KV
state frequently enough to restore arbitrary playback boundaries consumes memory
bandwidth and capacity that scale with sequence length and batch size, and
invalidating a shared prefix can disturb batching for co-scheduled requests. In
runtimes whose cache is a single continuous sequence, removing an unheard suffix
can degrade into a heavier re-prefill, erasing the latency advantage that
generating ahead of playback was meant to provide.

\noindent\textbf{Generality.}\quad Finally, item truncation is tied to a provider-specific
conversation-item abstraction, while cache rewinding is tied to one model
architecture and inference engine. Both approaches must be re-implemented and
re-validated across runtimes; neither naturally covers black-box voice models
that expose no truncation handle or cascaded VAD--ASR--LLM--TTS pipelines whose
stateless per-turn calls carry no persistent cache to rewind. In short, these
mechanisms can be useful adapters for particular systems, but they bind
playback grounding either to a provider API or to model internals while still
depending, for semantic restoration, on a playout-to-dialogue mapping they do
not supply by themselves.

%% file: tex/algo_design.tex
\section{The PACE Framework}\label{sec:pace}

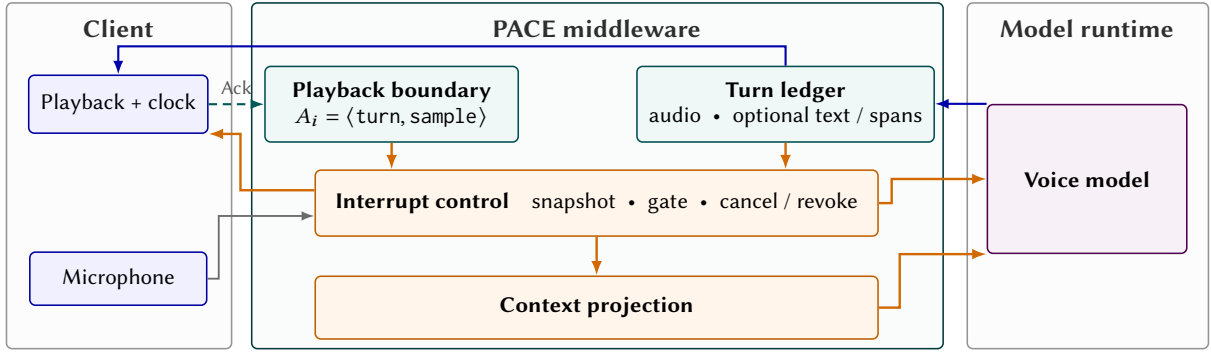
\begin{figure*}[t]
  \centering
  \resizebox{0.90\textwidth}{!}{\input{fig/pace-architecture}}
  \caption{PACE architecture and data flow. The upper path tracks streamed
  output and client playback acknowledgments to maintain a turn-local playback
  boundary. Upon an interruption, the lower path snapshots this boundary,
  cancels or revokes stale output, gates new user audio, and projects the
  grounded context into the model runtime.}
  \Description{Three-lane architecture diagram with a client on the left, PACE
  middleware in the center, and a model runtime on the right. PACE tracks model
  output and playback progress, controls interruption, and projects the heard
  context back into the model runtime.}
  \label{fig:pace-architecture}
  \vspace{-0.2cm}
\end{figure*}

PACE is a middleware layer placed between the transport stack and the model
runtime. It does not prescribe a specific LLM, ASR, TTS, or transport protocol.
Instead, it defines a small set of state objects and events that let the
runtime align its conversational context with the user's playback boundary.
Fig.~\ref{fig:pace-architecture} provides an architectural overview of PACE.

\subsection{Design Principles}\label{sec:principles}

\noindent\textbf{Separate generation from grounding.}\quad PACE records generated output
immediately, but marks samples as playback-grounded only after a client
acknowledgment passes them. When audio-to-content alignment is available, a
semantic unit becomes eligible for commitment only after playback passes its
finalized audio endpoint. This preserves low-latency streaming while avoiding
premature context commitment.

\noindent\textbf{Use turn-local coordinates.}\quad Playback samples and text offsets are
reported relative to a stable assistant turn identifier. A global sample count
is fragile because buffers can be reset, turns can overlap, and transport
streams can be reordered. Turn-local coordinates make acknowledgments
composable across WebTransport, WebSocket, WebRTC, and SIP.

\noindent\textbf{Project to the available runtime.}\quad PACE does not require every backend
to expose the same state-control interface. A cascaded text LLM can be
controlled by rewriting prompt history. A black-box realtime voice API can be
controlled through audio re-injection when truncate events are unavailable, or
through truncate and cancel events when available. A self-hosted
speech-to-speech model can expose stronger checkpoint, rewind, or mask
operations.

\noindent\textbf{Gate irreversible actions.}\quad It is acceptable for speculative speech to
be generated ahead of playback. It is not acceptable for irreversible tool
calls to commit based on proposals that were never grounded. PACE therefore
separates response generation from action commitment.

\subsection{Core Abstractions}\label{sec:core-state}\label{sec:playback-clock}

PACE introduces two primary state objects: the \emph{OutputTurnLedger}, which
lives on the server, and the \emph{PlaybackAck}, which flows from client to
server.

\noindent\textbf{Turn and OutputTurnLedger.}\quad
A \emph{turn} in PACE is not the model's output content itself, but the
lifecycle process of one assistant response---spanning generation, network
delivery, client playback, and availability for user perception (or
revocation). Each
time the model starts a new response (signaled by a
\texttt{response.created} event), PACE opens a new turn and assigns it a
sequential identifier~$i$. The
\emph{OutputTurnLedger} maintains one entry~$T_i$ per turn, recording both
the content produced and the delivery progress observed so far:
\begin{itemize}[leftmargin=*]
  \item \texttt{turn\_id}~($= i$): a monotonically increasing identifier
  assigned when the model begins a new response. It serves as the primary key
  that binds all related artifacts---downlink audio deltas, client playback
  reports, and projections---to a single response episode;
  \item \texttt{audio\_copy}: a server-side retained copy of the audio
  chunks delivered to the client for this turn, bounded to the most recent
  $W_{\max}$ seconds. Without this copy, the played prefix could not be
  recovered for context re-injection after an interruption;
  \item \texttt{text\_copy} (optional): the textual content of the assistant's
  response for this turn. While \texttt{audio\_copy} preserves what the
  response \emph{sounded like}, \texttt{text\_copy} records what it
  \emph{said}---enabling risk evaluation and history rewriting and, when paired
  with audio alignment, semantic-level boundary placement;
  \item \texttt{semantic\_spans} (optional): structured annotations over the
  turn's content. Each span records a turn-local audio interval
  $[b_j,e_j]$, a \texttt{finalized} flag, and a dialogue role---e.g., plain
  statement, proposal, question, tool-intent, or confirmation request. A span
  may additionally carry a text or semantic-token range, a runtime checkpoint,
  and alignment confidence;
  \item \texttt{played\_samples}: the sample offset of the \emph{playback
  boundary}---how far into this turn's audio the client has rendered to the
  speaker, updated on each PlaybackAck. Content before this offset is
  playback-grounded; content after it is speculative;
  \item \texttt{state}: the lifecycle phase of this turn---\emph{generating}
  $\rightarrow$ \emph{sent} $\rightarrow$ \emph{playing} $\rightarrow$
  \emph{playback-complete}, or \emph{revoked} if interrupted before completion.
\end{itemize}
For cascaded systems, \texttt{text\_copy} is the text LLM output and
\texttt{audio\_copy} is filled by TTS. For native speech-to-speech models,
\texttt{text\_copy} is often unavailable; when the model does expose an aligned
view, it may take the form of a streaming transcript or a semantic-token
sequence. Such a view supports semantic commitment only if its finalized units
are also mapped to output sample positions. A latent token sequence or inner
monologue without this mapping is not sufficient.

\noindent\textbf{PlaybackAck.}\quad
The client continuously reports its playback progress via a
\emph{PlaybackAck}:
\begin{equation}
  A = \langle \texttt{audio\_turn\_id},\; \texttt{played\_samples} \rangle .
  \label{eq:playback-ack}
\end{equation}
The field \texttt{audio\_turn\_id} identifies \emph{which} turn is
currently being rendered by the client speaker, while
\texttt{played\_samples} indicates \emph{how far} into that turn the client
has rendered, expressed as a cumulative sample count. Together, the two
fields identify the active turn's \emph{playback boundary}.

\noindent\textbf{Onset anchoring.}\quad
Because the client playback clock is the sole authority on playout timing,
and each uplink microphone tick co-transmits the audio frame and the
instantaneous \texttt{played\_samples} at the moment of capture, the
server can anchor the playback boundary directly to the user's
\emph{speech onset}. Upon detecting that the user has begun speaking, the
server-side VAD locates the onset tick and reads its
\texttt{played\_samples} as the cut-off line: all audio preceding this point
had reached playback when the user began speaking, whereas any audio that
continues to play out after this instant falls outside the conversational basis
for the interruption.

\subsection{Playback-Boundary Projection}
\label{sec:playback-boundary-projection}

PACE always exposes a physical \emph{PlaybackBoundary}:
\begin{equation}
  B_i^{p} = \langle \texttt{audio\_turn\_id}=i,\;
  \texttt{played\_samples}=n \rangle .
\end{equation}
When aligned semantic spans are available, PACE may additionally derive a
\emph{semantic commit boundary}. Let $e_j$ be the aligned audio endpoint of a
finalized span $s_j$, and let $\epsilon$ bound acknowledgment and alignment
uncertainty. Then
\begin{equation}
  B_i^{s}(n) = \max\bigl(\{e_j \mid
  \operatorname{finalized}(s_j) \land e_j + \epsilon \leq n\}\cup\{0\}\bigr).
  \label{eq:semantic-commit-boundary}
\end{equation}
The runtime or online segmenter sets \texttt{finalized} only after declaring a
span closed; provisional streaming hypotheses do not qualify. If the runtime
and middleware provide no audio-to-content alignment, $B_i^{s}(n)=\bot$: PACE
does not infer a semantic boundary from the waveform or from elapsed time alone.

The two boundaries serve different purposes. $B_i^{p}$ controls media
revocation and preserves the exact played audio prefix, including a partial
utterance that may have prompted the interruption. $B_i^{s}$ controls which
complete propositions may enter canonical text history or satisfy an action
precondition. A third, runtime-specific \emph{restoration boundary} identifies
the latest checkpoint, token position, or conversation item to which internal
state can actually be restored. It may be coarser than either boundary or may
not exist. When user activity begins, PACE snapshots $B_i^{p}$ and uses the
strongest projection supported by the runtime. Projection is therefore not
synonymous with state rewind: it may rewrite history, truncate a conversation
item, re-inject played audio, or restore native model state.

\noindent\textbf{Cascaded adapter.}\quad A cascaded voice agent typically calls a stateless
chat-completion API on each turn. For this class, projection is implemented by
rewriting history. TTS timestamps or forced alignment map $B_i^{p}$ to a text
offset; if finalized spans are available, the canonical history is truncated
at $B_i^{s}$. If no reliable alignment exists, PACE does not estimate a text
offset from character count or elapsed time: it either removes the incomplete
assistant turn or uses the audio re-injection adapter. The next LLM call
receives a short system context such as: ``The previous assistant response was
interrupted; only the following prefix was played.'' This adapter requires no
model changes and is practical for many VAD--ASR--LLM--TTS systems.

\noindent\textbf{Audio re-injection adapter.}\quad When the model runtime does not expose
item-level truncation or history editing (e.g., a black-box realtime voice API
accessed only through audio input/output buffers), PACE projects the playback
boundary by re-injecting the played audio suffix as model input. The procedure
operates in three steps:

\begin{enumerate}[leftmargin=*]
  \item \textit{Extract played context.}\quad PACE first reads the playback position
  reported by the client: \texttt{played\_samples}. It then uses this value and
  the turn-indexed downlink buffer to slice the most recent $L$ seconds of
  played audio (the \emph{lookback window}).
  Let $n$ denote \texttt{played\_samples} and $f_p$ the playback sample rate.
  The slice boundary is $[s,\; n]$, where $s=\max(0,\; n - L \cdot f_p)$.

  \item \textit{Resample and concatenate the instruction delimiter.}\quad The extracted segment
  (at the playback rate $f_p$) is resampled to the model's upstream
  rate $f_u$ via linear interpolation. A pre-synthesized audio
  prompt $P_{\text{delim}}$ is appended to mark the preceding segment as
  assistant output already heard by the user and to separate it from the
  post-interrupt request. In our deployment, $P_{\text{delim}}$ is the TTS
  rendering of the fixed phrase ``That was what I heard before interrupting.
  Now, here is what I want to say:'', synthesized once at server startup and
  cached on disk as raw PCM.

  \item \textit{Inject as model input.}\quad PACE sends the concatenated audio to the
  model's input buffer: first the heard assistant context, then the instruction
  delimiter, and finally the user speech captured after the interruption. The
  delimiter directs the model to treat the preceding audio as previously heard
  context and to answer the request that follows. This sequence grounds the
  model's next response at the playback boundary.
\end{enumerate}

This audio-only fallback requires an audio input path and a way to cancel or
suppress stale output, but makes no assumption about text, conversation-item
APIs, or KV-cache accessibility. It preserves the physical played prefix; it
does not by itself establish a proposition-level boundary or exact internal
state rollback.

\noindent\textbf{Native speech-to-speech adapter.}\quad Native runtimes expose three
useful capability levels. A cooperative runtime may emit finalized transcript
or semantic-token spans together with output-sample endpoints and restoration
checkpoints; PACE then computes $B_i^{s}$ and restores the latest compatible
checkpoint. When only output audio is exposed, an external streaming ASR and
online span segmenter may populate approximate \texttt{semantic\_spans}; their
alignment error must be reflected in $\epsilon$ and evaluated explicitly. A
fully black-box runtime with neither form of alignment has
$B_i^{s}=\bot$. PACE then uses $B_i^{p}$ for revocation and audio re-injection,
and treats semantic preconditions as ungrounded until an explicit confirmation.

\subsection{Triggering and Interruption Handling}
\label{sec:interrupt-handling}\label{sec:cancel}
\label{sec:forward-gate}\label{sec:trigger}

\noindent\textbf{Trigger policy.}\quad
PACE applies grounding projection when it observes a \emph{risk event}. Common
risk events include confirmed user interruption, explicit push-to-talk down,
manual interrupt, new response replacing an old one, transport reconnection,
and tool commit. An acoustic speech-start signal is treated as a candidate
risk event rather than an unconditional projection trigger: short backchannels,
echo leakage, and false positives may update the playback boundary without
canceling the assistant's current output.

A more selective policy, suitable for runtimes with richer semantic
information, also considers semantic risk:
\begin{itemize}[leftmargin=*]
  \item the unheard suffix contains a proposal, question, correction, or tool
  precondition;
  \item the user utterance contains a deictic phrase such as ``that'', ``last
  one'', ``OK'', ``repeat'', or ``why'';
  \item the system is about to execute an external action whose proposal span
  is not grounded.
\end{itemize}

\noindent\textbf{In-flight cancellation.}\quad
Projecting the playback boundary into the model-facing context does not remove
output already in flight.
By the time the server processes an interruption, audio from the interrupted
response may still be queued in the transport pipeline or the client playback
buffer. PACE therefore cancels the response along two paths: it stops and
suppresses further output at the source, and revokes buffered output at the
sink.
\begin{enumerate}[leftmargin=*]
  \item \textit{Upstream cancellation.}\quad PACE issues
  \texttt{response.cancel} to stop further generation. At the same time, the
  middleware raises a \emph{suppression
  flag}, preventing any remaining audio or text deltas from the interrupted
  response from reaching the client.
  \item \textit{Downstream revocation.}\quad PACE also sends the client a
  Revoke event:
  \begin{equation}
    R = \langle \texttt{audio\_turn\_id},\; \texttt{after\_sample},\;
    \texttt{reason}\rangle .
  \end{equation}
  Here, \texttt{audio\_turn\_id} identifies the interrupted response and
  \texttt{after\_sample} marks the last sample that may be retained. Upon
  receipt, the client discards all scheduled or buffered audio beyond this
  boundary, preserving only the prefix already rendered.
\end{enumerate}

PACE keeps the suppression flag active until the next
\texttt{response.created} event. This blocks late deltas from the interrupted
response, which black-box runtimes may emit even after cancellation.

\noindent\textbf{Forward gate and ordering.}\quad
Context re-injection must be sequenced correctly relative to user speech: the
model should see the injected heard-context audio \emph{before} the
post-interrupt user utterance, never after. Without explicit ordering, user
speech ticks that were in-flight or queued during the interrupt could leak
ahead of the injected context, causing the model to hear user speech without
the grounding prefix.

PACE enforces ordering through a \emph{forward gate} mechanism:
\begin{enumerate}[leftmargin=*]
  \item When an interrupt event is detected, the gate is closed
  \emph{synchronously}, before any asynchronous operation begins. This blocks
  subsequent uplink ticks from reaching the model backend.
  \item The middleware acquires a send-serialization lock, executes
  \texttt{response.cancel}, builds the re-injection payload, and transmits
  it to the model.
  \item Any ticks that passed the gate before closure but were waiting for
  the send lock are \emph{held} rather than dropped; they are flushed in
  order after the injected context, preserving the user's post-interrupt speech.
  \item The gate is re-opened only after injection completes, allowing normal
  tick forwarding to resume.
\end{enumerate}
This guarantees the model receives the sequence: \emph{last pre-interrupt
tick} $\rightarrow$ \emph{cancel} $\rightarrow$ \emph{injected context
(played audio + delimiter)} $\rightarrow$ \emph{held ticks} $\rightarrow$
\emph{new user speech}.

%% file: fig/pace-architecture.tex
\begin{tikzpicture}[
  x=1cm,
  y=1cm,
  font=\sffamily\footnotesize,
  >=Latex,
  lane/.style={draw=black!42, rounded corners=2pt, line width=0.55pt,
               fill=black!1},
  module/.style={draw=black!48, rounded corners=2pt, line width=0.55pt,
                 fill=white, align=center, inner sep=4.2pt},
  clientmod/.style={module, draw=blue!62!black, fill=blue!6},
  pacemod/.style={module, draw=teal!62!black, fill=teal!6},
  repairmod/.style={module, draw=orange!75!black, fill=orange!8},
  runtimemod/.style={module, draw=violet!62!black, fill=violet!6},
  header/.style={font=\sffamily\bfseries\small, text=black!82},
  sub/.style={font=\sffamily\scriptsize, text=black!62, align=center},
  data/.style={draw=blue!68!black, line width=0.75pt, -{Latex[length=1.7mm]}},
  feedback/.style={draw=teal!70!black, densely dashed, line width=0.75pt,
                   -{Latex[length=1.7mm]}},
  interrupt/.style={draw=orange!82!black, line width=0.85pt,
                    -{Latex[length=1.8mm]}},
  neutral/.style={draw=black!58, line width=0.65pt,
                  -{Latex[length=1.6mm]}}
]
  \node[lane, minimum width=2.80cm, minimum height=4.30cm, anchor=south west]
    (clientlane) at (0,0) {};
  \node[lane, draw=teal!45!black, fill=teal!2, minimum width=8.60cm,
        minimum height=4.30cm, anchor=south west] (pacelane) at (3.05,0) {};
  \node[lane, minimum width=3.00cm, minimum height=4.30cm, anchor=south west]
    (runtimelane) at (11.95,0) {};

  \node[header, anchor=north] at (1.40,4.23) {Client};
  \node[header, anchor=north] at (7.35,4.23) {PACE middleware};
  \node[header, anchor=north] at (13.45,4.23) {Model runtime};

  \node[clientmod, minimum width=2.20cm, minimum height=0.72cm]
    (playback) at (1.40,3.05)
    {Playback + clock};
  \node[clientmod, minimum width=2.20cm, minimum height=0.66cm]
    (capture) at (1.40,0.88) {Microphone};

  \node[pacemod, minimum width=3.15cm, minimum height=0.94cm]
    (tracker) at (4.80,3.05)
    {\textbf{Playback boundary}\\[-1pt]
     $A_i=\langle\mathtt{turn},\mathtt{sample}\rangle$};
  \node[pacemod, minimum width=3.30cm, minimum height=0.94cm]
    (ledger) at (9.70,3.05)
    {\textbf{Turn ledger}\\[-1pt]
     audio \;\textbullet\; optional text / spans};

  \node[repairmod, minimum width=7.00cm, minimum height=0.82cm]
    (coordinator) at (7.35,1.82)
    {\textbf{Interrupt control}\quad
     snapshot \;\textbullet\; gate \;\textbullet\; cancel / revoke};

  \node[repairmod, minimum width=7.00cm, minimum height=0.78cm]
    (projection) at (7.35,0.52) {\textbf{Context projection}};

  \node[runtimemod, minimum width=2.48cm, minimum height=1.85cm,
        font=\sffamily\bfseries\footnotesize]
    (runtime) at (13.45,2.12) {Voice model};

  \draw[data] (runtime.north west) --
    (ledger.east |- runtime.north west);
  \draw[data] (ledger.north) -- ++(0,0.24) -| (playback.north);

  \draw[feedback] (playback.east) -- node[above, sub, text=black!66]
    {Ack} (tracker.west);
  \draw[neutral] (capture.east) -- ++(0.14,0) |-
    ([yshift=-0.16cm]coordinator.west);

  \draw[interrupt] (tracker.south) -- (coordinator.north -| tracker.south);
  \draw[interrupt] (ledger.south) -- (coordinator.north -| ledger.south);
  \draw[interrupt] (coordinator.south) -- (projection.north);
  \draw[interrupt] (projection.east) -- ++(0.26,0) |- (runtime.south west);
  \draw[interrupt] (coordinator.east) -- ++(0.16,0) |- (runtime.west);
  \draw[interrupt] ([yshift=0.16cm]coordinator.west) -- (2.92,1.98) |-
    (playback.south east);
\end{tikzpicture}

%% file: tex/implementation.tex
\section{Implementation}\label{sec:implementation}

We implemented PACE in a browser-based realtime voice assistant. The prototype
consists of a JavaScript client, an asynchronous Python middleware server, and
an adapter to the DashScope Qwen-Audio-Realtime API. PACE requires no changes
to the model service: playback tracking runs at the client, while the ledger,
ordering, and context projection run in the middleware. The evaluation uses
this same end-to-end path; the benchmark runner replaces only the live
microphone source with a prerecorded user-input trajectory.

\subsection{Prototype and Playback Tracking}\label{sec:prototype-data-path}

The browser sends float32 microphone audio in 100-ms ticks. WebTransport is the
primary client--server transport, using unidirectional streams for media ticks
and a long-lived bidirectional stream for control messages; WebSocket provides
a compatibility fallback. The server reorders WebTransport ticks by a 16-bit
sequence number, waiting at most 50~ms for a missing tick. Interrupt-marked
ticks bypass this queue and are handled immediately.

Assistant audio is scheduled with the Web Audio clock. For each assistant
turn, the client counts fully rendered chunks and the elapsed portion of the
active chunk. Every uplink tick header carries the current 16-bit
\texttt{audio\_turn\_id} and 32-bit \texttt{played\_samples} offset at
24~kHz, continuously reporting the latest playback progress. At interruption,
the server reads the values carried by the speech-onset tick, and the client
then stops active and queued audio from that turn. These fields extend the
existing tick header to 33~bytes. Each downlink delta carries a 27-byte
header containing its delta ID, sample rate, audio length, end-of-turn flag,
audio-turn ID, source tick ID, and server timestamp. 

\subsection{Server-Side Execution}\label{sec:server-execution}

The middleware copies each model audio delta into a turn-indexed buffer before
forwarding it to the client. It retains at most eight turn entries, together
with a 10-s live fallback buffer and the previous response buffer. This layout
allows an interrupt report to select the turn actually being played even when
the model has already emitted \texttt{response.created} for a newer turn.

The interrupt fast path closes the forward gate, increments an interruption
generation counter, and copies the live and previous-response buffers before
the first asynchronous yield. The interrupt handler then acquires the same
send lock used by normal model-input writes. Under this lock, it issues
\texttt{response.cancel}, injects the repaired context, and releases held
microphone ticks in order. A \texttt{finally} path always reopens the gate,
while output suppression remains active until the next
\texttt{response.created} event. These implementation-level safeguards cover
both ticks that arrive after gate closure and asynchronous senders that had
already passed the gate.

For the black-box model adapter, the middleware extracts at most 5~s of played
audio, resamples it from 24-kHz float32 to 16-kHz PCM16, and appends the spoken
instruction delimiter $P_{\text{delim}}$ described in
Section~\ref{sec:playback-boundary-projection}. The delimiter is synthesized once
with the assistant voice and cached as raw PCM. The played context and delimiter
are appended to the model's audio-input buffer in chunks of at most 192,000 raw
bytes before held user audio is released.

The prototype therefore implements the physical boundary $B_i^{p}$ and the
audio-only projection path. The model backend adapter does not expose an aligned
transcript or restorable semantic checkpoint, so the prototype does not
populate \texttt{semantic\_spans} or compute $B_i^{s}$. Its guarantee is that
unplayed audio remains externally revocable and that the played prefix is
presented before post-interruption user speech; it does not claim exact
proposition-level commitment or internal-state equivalence.

%% file: tex/dataset.tex
\section{GCM-Bench: A Dataset for Generative Context Mis-anchoring}
\label{sec:dataset}

This section presents GCM-Bench, a controlled benchmark of 108 interruption
cases designed to measure GCM. Its central task is playback-relative referent
anchoring: after an interruption, the system must ground its response in the
content item at the playback boundary rather than in unheard, model-internal
continuation. We first motivate the dataset and explain why existing
interruption benchmarks do not capture GCM
(Section~\ref{sec:benchmark-dataset}), then describe the factorial
construction of the 108 cases (Section~\ref{sec:gcm-scenarios}), and finally
detail the end-to-end measurement protocol and the Referent Anchoring Accuracy
metric (Section~\ref{sec:evaluation-methodology}).

\subsection{Dataset Objective}\label{sec:benchmark-dataset}

Existing interruption benchmarks evaluate whether a voice agent yields the floor
and whether its subsequent response is relevant to the user's new request.
These criteria, however, do not test GCM. The user-interruption track of
Full-Duplex-Bench (FDB)~\cite{fullduplexbench2025}, for instance, reports
turn-obedience rate (TOR), response quality, and latency, yet it does not
identify which portion of the interrupted response the model treats as the
referent. A model may therefore elaborate an internally generated fifth list
item and still receive a high relevance score even though the user had only
heard the third.

To fill this gap, we construct GCM-Bench, a new controlled benchmark dataset
designed specifically to measure GCM. It defines a \emph{playback-relative}
referent-anchoring task and contains 108 controlled interruption cases. Its
primary metric is \emph{Referent Anchoring Accuracy} (RAA)---whether a
post-interruption response is anchored to the content item at the playback
boundary rather than to an unheard, model-internal item.

We use 200 official FDB v1 user-interruption samples in a separate
compatibility study. These external samples are not part of GCM-Bench. The FDB
study tests whether context repair degrades ordinary topic-switching
interruptions and provides an independent latency check.

\subsection{GCM-Bench Construction}\label{sec:gcm-scenarios}

We construct GCM-Bench as the full Cartesian product of 12 listing scenarios,
three interruption delays, and three referential operations, yielding
$12\times3\times3=108$ samples (Table~\ref{tab:raa-composition}). The scenarios
span three discourse structures: four concept lists (e.g., programming
languages), four procedural lists (e.g., preparing scrambled eggs with
tomatoes), and four preparation or requirement lists (e.g., traveling abroad).
We choose listing tasks because they produce a sequence of discrete,
identifiable items whose boundaries are externally verifiable---an evaluator
can determine unambiguously which item occupies the playback boundary---while
still preserving natural variation in item length and discourse dependency.

\begin{table}[t]
  \caption{GCM-Bench factorial composition. Crossing 12 scenarios (four per
  class), three delays, and three referential operations yields 108 cases;
  \#/level denotes the cases at each listed level.}
  \label{tab:raa-composition}
  \vspace{-0.5em}
  \centering
  \small
  \begin{tabular}{@{}lcc@{}}
    \toprule
    Dimension & Levels & \#/level \\
    \midrule
    Scenario class & concept, procedure, condition & 36 \\
    Interrupt. delay & 12 s, 16 s, 20 s & 36 \\
    Referential operation & \emph{elaborate}, \emph{next}, \emph{repeat} & 36 \\
    \bottomrule
  \end{tabular}
\end{table}

The three operations exercise distinct uses of the playback context.
\emph{Elaborate} (``Tell me more about this one'') requires expanding the item
at the boundary; \emph{next} (``Tell me about the next one'') requires advancing
relative to that item; and \emph{repeat} (``I didn't catch that, say it again'')
requires reproducing it verbatim. Together, they probe whether the system can
locate, extend, and advance from the playback-grounded referent.

Interruptions occur at 12, 16, or 20~s after the end of the initial query,
placing the same referential request at progressively later stages of the
generated list. We arrived at these delays empirically: an initial 5--9~s range
frequently interrupted before the model had established a stable item,
causing the task to measure response startup latency rather than referent
anchoring.

We manually author the 12 source prompts and programmatically generate all 108
GCM-Bench cases. User utterances are synthesized with DashScope
Qwen3-TTS-Flash using the Cherry voice, resampled to 16~kHz, stripped of
trailing silence, padded with 0.3~s of silence, and peak-normalized. Each case
packages a reproducible user-input trajectory that concatenates the initial
query, the designated delay, the interruption utterance, and 15~s of trailing
silence. The construction process records the exact interruption onset and
endpoint consumed by the streaming runner and evaluator. The trajectory fixes
only user-side input and timing; assistant generation, network delivery,
browser playback, and interruption repair execute online in every trial.

\subsection{Measurement Protocol}\label{sec:evaluation-methodology}

GCM-Bench evaluation proceeds in three stages: streaming execution,
transcription, and automated judging.

\noindent\textbf{Streaming execution.}\quad Each sample runs through the
deployed end-to-end system in a fresh model session.
The runner feeds the 16-kHz user-input trajectory into the same uplink path used
by live microphone audio, at wall-clock rate. Returned assistant audio is
scheduled through the browser's Web Audio playback path, whose clock produces
the PlaybackAck consumed by PACE. A synchronized recorder captures the live
model output while preserving its temporal correspondence with user input and
actual client playback. The benchmark supplies neither assistant output nor a
playback boundary and does not simulate playback.

\noindent\textbf{Transcription.}\quad We transcribe the full assistant audio
with the Qwen3-ASR service, retaining word-level timestamps. The playback
boundary captured at interruption partitions each transcript into
\emph{before-text} (the assistant speech already played when the interruption
began) and \emph{after-text} (the response generated after the interruption).

\noindent\textbf{Automated judging.}\quad We report TOR separately from
anchoring quality. Under the evaluation protocol, a response shorter than 1~s
with at most three words is classified as no effective response. For every
effective response, ChatGPT 5.5 receives the before-text, after-text,
interruption utterance, and operation type. The judge returns a binary decision
with an explanation indicating whether the after-text elaborates, advances from,
or repeats the referent at the end of the played prefix. Conditional on an
effective response, RAA is
\begin{equation}
  \operatorname{RAA}=\frac{1}{N_{\mathrm{resp}}}
  \sum_{k=1}^{N_{\mathrm{resp}}}
  \mathbb{1}[\hat y_k=\mathrm{correct}].
  \label{eq:raa}
\end{equation}
Crucially, the judge sees the model's actual output before interruption, not a
canonical item list. This design accommodates the fact that independent realtime
runs may enumerate items at different rates or with different wording. Reporting
TOR alongside RAA prevents a system from inflating conditional accuracy by
simply refusing to respond. In our benchmark, however, every system achieves
$\mathrm{TOR}=100\%$ ($N_{\mathrm{resp}}=N$), so the conditioning is vacuous
and RAA reduces to unconditional accuracy.

Posterior judging is reliable for \emph{elaborate} and \emph{repeat}: the played item alone
suffices to determine whether the response uses the correct referent. For \emph{next},
however, the method is weaker. The judge observes the played prefix but not the
generated-yet-unplayed continuation, and therefore cannot always verify the
identity and order of the true next item. It may accept a new, non-repeated item
even when the model has advanced from an internal item beyond the playback
boundary. We accordingly report \emph{next} separately and treat \emph{elaborate} as the
cleanest direct measure of GCM repair.

For the separate FDB compatibility study, we retain the original v1 evaluation
protocol unmodified. We crop output from the end of the interruption,
transcribe it, and correct the first-word timestamp using audio energy. We then
report TOR, the original 0--5 LLM-judge response score, and
interruption-to-response latency. Because these external FDB samples
predominantly contain topic switches (e.g., ``Actually, can we talk about
movies instead?'') rather than referential requests, they are not used to
estimate RAA.

%% file: tex/evaluation.tex
\section{Evaluation}\label{sec:eval}

This section evaluates PACE on the deployed end-to-end system, organized
around three research questions on anchoring, robustness, and
compatibility. On the 108 paired GCM-Bench cases, playback-context projection
raises Referent Anchoring Accuracy from 25.0\% to 96.3\%, and a separate study
on 200 FDB v1 samples confirms that ordinary interruption handling is
preserved at a modest repair cost.

\subsection{Experimental Setup}\label{sec:methodology}

Our evaluation addresses three questions:
\begin{itemize}[leftmargin=*]
  \item \textbf{RQ1: Anchoring.} Does playback-context projection reduce
  referent mis-anchoring relative to cancellation alone?
  \item \textbf{RQ2: Robustness.} Does the effect persist across interruption
  times, referential operations, and discourse structures, and how sensitive is
  it to the delimiter and lookback-window design?
  \item \textbf{RQ3: Compatibility and cost.} Does projection preserve
  interruption quality, and what latency does interruption repair add?
\end{itemize}

We evaluate anchoring and robustness on the newly constructed 108-case
GCM-Bench dataset. Separately, we assess compatibility and latency on 200
official FDB v1 user-interruption samples; these external samples are not part
of GCM-Bench.

All trials exercise the deployed PACE path end to end. GCM-Bench supplies only
the prerecorded user-input trajectory: assistant output is generated online,
scheduled by the browser for actual playback, and tracked by client PlaybackAck
events. The middleware consumes those acknowledgments to update the
OutputTurnLedger, snapshot the interruption boundary, revoke queued audio, and
order cancellation and context projection. The benchmark runner neither
precomputes nor injects a playback position.

Each sample is evaluated under a \emph{cancellation-only baseline} and \emph{PACE}.
The baseline cancels the active response and forwards the interruption without
context repair. PACE additionally snapshots the client-reported playback
boundary and extracts the latest 5~s of played assistant audio from the
server-side turn-indexed output buffer. Let $H$ denote this segment and $U$ the
post-interrupt user audio. Following the notation of
Section~\ref{sec:playback-boundary-projection}, the deployed configuration uses
$P_{\text{delim}}=\mathrm{P1}$; P2 and P3 are evaluated only as ablations:
\begin{itemize}[leftmargin=*,nosep]
  \item \textbf{P1 (natural statement):} $H \rightarrow$ ``That was what I
  heard before interrupting. Now, here is what I want to say:'' $\rightarrow U$.
  \item \textbf{P2 (direct handoff):} $H \rightarrow$ ``That was what I heard.
  Now please respond to my following request instead:'' $\rightarrow U$.
  \item \textbf{P3 (wrapped context):} ``I heard you say the following:''
  $\rightarrow H \rightarrow$ ``Now please respond to my following request
  instead:'' $\rightarrow U$.
\end{itemize}
Each instruction is synthesized in the assistant voice and injected through the
model's audio-input interface in the order shown. This audio-only condition
instantiates $B_i^p$ without semantic spans or model-state checkpoints.

Both conditions use the same user-input trajectories, interruption timestamps,
realtime model, and inference configuration. Each sample is run independently
in each condition, and the judge conditions on the output actually produced in
that run. We report percentage-point (pp) changes for RAA and apply McNemar's
test with continuity correction to the 108 paired GCM-Bench outcomes. We
additionally ablate three delimiter designs at a fixed 5-s window and three
lookback windows with the selected delimiter. Cochran's Q tests compare each set
of three paired conditions, followed by exact pairwise McNemar tests with Holm
correction. The 2.5- and 10-s lookback conditions were collected in a later
realtime batch than the 5-s condition; we therefore treat their near-tied
results cautiously.
For the paired FDB study, we report the score difference descriptively and use
paired bootstrap intervals and a Wilcoxon signed-rank test for latency.

\subsection{RQ1: Referent Anchoring}\label{sec:eval-results}

Both conditions produce an effective response for all 108 GCM-Bench samples
(TOR $=100\%$). The baseline anchors 27 responses correctly (25.0\%), whereas
PACE anchors 104 (96.3\%), an improvement of 71.3 pp
(Table~\ref{tab:raa-main}). The result isolates a failure that conventional
interruption metrics miss: both systems respond, but only the playback-aligned
condition usually resolves what the user is referring to.

\begin{table}[t]
  \caption{Referent Anchoring Accuracy (RAA) for the cancellation-only baseline
  and PACE on 108 paired GCM-Bench cases, overall and by operation. Both achieve
  100\% TOR.}
  \label{tab:raa-main}
  \vspace{-0.5em}
  \centering
  \small
  \begin{tabular}{@{}lrrr@{}}
    \toprule
    Operation & Baseline & PACE & $\Delta$ \\
    \midrule
    \emph{elaborate} & 1/36 (2.78\%) & 34/36 (94.44\%) & +91.67 pp \\
    \emph{next}      & 22/36 (61.11\%) & 35/36 (97.22\%) & +36.11 pp \\
    \emph{repeat}    & 4/36 (11.11\%) & 35/36 (97.22\%) & +86.11 pp \\
    \midrule
    Overall   & 27/108 (25.0\%) & 104/108 (96.3\%) & +71.3 pp \\
    \bottomrule
  \end{tabular}
\end{table}

The paired outcomes comprise 26 samples correct under both conditions, one
baseline-only success, 78 PACE-only successes, and three failures under both.
McNemar's test with continuity correction yields $p<0.001$. Thus, the
improvement is not explained by a small number of cases: PACE repairs 78
baseline failures while introducing one regression, a 78:1
improvement-to-regression ratio.

The operation breakdown shows where the repair pipeline matters. \emph{Elaborate}
improves from 2.78\% to 94.44\%; without repaired playback context, the model
almost never resolves ``this one.'' \emph{Repeat} improves from 11.11\% to 97.22\%,
and \emph{next} from 61.11\% to 97.22\%, though its baseline is likely inflated by the
judging limitation in Section~\ref{sec:evaluation-methodology}. Accordingly,
\emph{elaborate} provides the cleanest evidence for playback-relative anchoring.

A representative case illustrates the difference. When the played output ends
while introducing bananas and the user asks ``Tell me more about this one,''
the baseline asks which fruit the user means. PACE instead identifies bananas
and elaborates on their nutritional properties. Both responses are fluent; the
difference is whether the referent is grounded at the user's playback boundary.

\subsection{RQ2: Robustness Across Conditions}\label{sec:robustness}

PACE exceeds 91\% RAA at every tested interruption time
(Table~\ref{tab:raa-delay-category}). It reaches 91.7\%, 100.0\%, and 97.2\% at
12, 16, and 20~s, respectively, compared with baseline accuracies of 33.3\%,
19.4\%, and 22.2\%. We do not observe a monotonic decline with delay: PACE
reaches 100\% at 16~s and 97.2\% at 20~s. This pattern suggests that
performance depends on which discourse item falls within the lookback, not on
elapsed conversation time alone.

\begin{table}[t]
  \caption{Referent Anchoring Accuracy (RAA) for the cancellation-only baseline
  and PACE on GCM-Bench, stratified by interruption delay and scenario class.}
  \label{tab:raa-delay-category}
  \vspace{-0.5em}
  \centering
  \small
  \begin{tabular}{@{}lrrr@{}}
    \toprule
    Group & Baseline & PACE & $\Delta$ \\
    \midrule
    Delay: 12 s & 12/36 (33.3\%) & 33/36 (91.7\%) & +58.3 pp \\
    Delay: 16 s & 7/36 (19.4\%) & 36/36 (100.0\%) & +80.6 pp \\
    Delay: 20 s & 8/36 (22.2\%) & 35/36 (97.2\%) & +75.0 pp \\
    \midrule
    Concept lists   & 12/36 (33.3\%) & 35/36 (97.2\%) & +63.9 pp \\
    Procedure lists & 2/36 (5.6\%)  & 33/36 (91.7\%) & +86.1 pp \\
    Condition lists & 13/36 (36.1\%) & 36/36 (100.0\%) & +63.9 pp \\
    \bottomrule
  \end{tabular}
\end{table}

The effect also persists across discourse structures. PACE obtains 97.2\% on
concept lists, 100.0\% on condition lists, and 91.7\% on procedural lists.
Concept items such as languages or fruits have distinct lexical boundaries and
are readily recovered from the injected audio. Procedural steps are more
causally and linguistically connected, making the current unit harder to
isolate. Nine scenarios reach 100\% RAA. The remaining errors concentrate in
two procedural scenarios---tomato-egg cooking and laundry---indicating broad,
but not uniform, robustness across the controlled conditions.

\noindent\textbf{Configuration ablations.}\quad
Table~\ref{tab:configuration-ablations} isolates the effects of two design
choices in the audio projection adapter: the instruction delimiter and the
playback lookback window. At a fixed 5-s lookback, P1 achieves the highest RAA
among the three tested delimiters. With P1 fixed, the 2.5- and 5-s windows yield
similarly high RAA, whereas the 10-s window substantially degrades anchoring.

\begin{table}[t]
  \caption{Ablations of the audio projection adapter on GCM-Bench. Each row
  evaluates one variant on all 108 cases: delimiter variants fix the lookback
  at 5~s, and lookback variants fix the delimiter at P1. The deployed
  configuration is P1 with a 5-s lookback.}
  \label{tab:configuration-ablations}
  \vspace{-0.5em}
  \centering
  \small
  \begin{tabular}{@{}llrr@{}}
    \toprule
    Ablation & Setting & Correct & RAA \\
    \midrule
    Delimiter & P1 (natural statement) & 104/108 & 96.30\% \\
              & P2 (direct handoff)   & 91/108  & 84.26\% \\
              & P3 (wrapped context)  & 99/108  & 91.67\% \\
    \midrule
    Lookback & 2.5 s & 103/108 & 95.37\% \\
             & 5 s   & 104/108 & 96.30\% \\
             & 10 s  & 71/108  & 65.74\% \\
    \bottomrule
  \end{tabular}
\end{table}

Delimiter choice has a significant overall effect (Cochran's $Q=8.897$,
$p=0.012$). After Holm correction, P1 significantly outperforms P2
($p=0.022$), but not P3. We therefore deploy P1 as the best-performing tested
delimiter without claiming a reliable advantage over P3. P2 performs
particularly poorly on \emph{next} (75.0\%), suggesting that its direct-handoff
wording may disrupt list continuation. Lookback length also has a significant
overall effect ($Q=51.561$, $p<0.001$). Both 2.5 and 5~s significantly
outperform 10~s after Holm correction ($p<10^{-6}$), whereas their 0.93-pp
difference is not significant ($p=1.000$). The 10-s window is particularly
harmful for \emph{repeat} (50.0\%), suggesting that the longer history introduces
competing candidate referents. Because the 2.5- and 10-s conditions were
collected in a later realtime batch, the near tie between 2.5 and 5~s should be
interpreted cautiously. We therefore retain 5~s as a conservative default and
regard 2.5~s as a viable lower-overhead alternative.

\subsection{RQ3: Compatibility and Repair Cost}\label{sec:compatibility}

Table~\ref{tab:fdb-results} reports the separate compatibility study on 200
official FDB v1 samples. Baseline and P1 run on the same 200 samples in the
same batch, with four concurrent workers and alternating execution order across
samples. Both conditions
respond to every interruption. The mean response-quality score changes only
from 4.975 for the baseline to 4.995 for P1. Mean response latency rises from
0.771 to 0.830~s, an additional 58.7~ms consistent with transmitting
and processing the injected audio context. This overhead occurs only after a
confirmed interruption, when PACE invokes context repair. A request issued
while the assistant is not playing follows the original, unchanged processing
path and therefore incurs no additional PACE latency.

\begin{table}[t]
  \caption{Compatibility results on 200 paired FDB v1 samples for the
  cancellation-only baseline and deployed PACE (P1 with a 5-s lookback).
  Quality is the original FDB v1 LLM-judge response-quality score.}
  \label{tab:fdb-results}
  \vspace{-0.5em}
  \centering
  \small
  \begin{tabular}{@{}lrrr@{}}
    \toprule
    Metric & Baseline & PACE & $\Delta$ \\
    \midrule
    TOR & 1.000 & 1.000 & 0.000 \\
    Quality score (0--5) & 4.975 & 4.995 & +0.020 \\
    Latency (s) & 0.771 & 0.830 & +0.059 \\
    \bottomrule
  \end{tabular}
\end{table}

The FDB result is a compatibility check rather than evidence of superiority on
general interruptions: FDB does not measure playback-relative anchoring, and we
report the score change descriptively. The latency increase is statistically
reliable (paired bootstrap 95\% CI $[24.8,96.4]$~ms; Wilcoxon
$p=6.66\times10^{-7}$), but remains below 0.1~s. Within that scope, PACE
preserves response rate and judged quality while introducing a measurable,
modest repair delay only on the interruption path.

\subsection{Error Analysis and Validity}\label{sec:error-analysis}

The four PACE errors fall into two recurring modes. One \emph{next} response (25\%)
repeats the current item rather than advancing. The remaining three errors
(75\%) occur in procedural scenarios and anchor to a neighboring sub-action
rather than the sub-action being played at interruption. These residual errors
suggest that PACE usually restores the relevant local context, while fine-grained
boundaries within a chain of closely related procedural actions remain difficult.

Three limitations bound the interpretation of our results. First, GCM-Bench
uses TTS-constructed user-input trajectories, listing tasks, one judge, and one
realtime model. The trajectories make user input and interruption timing
reproducible, but assistant generation, transport, browser playback, and repair
remain live; the study demonstrates the mechanism under controlled inputs
rather than estimating GCM prevalence in natural conversations. Second,
posterior judging can
overestimate \emph{next} accuracy because it lacks the unplayed item sequence. Third,
the experiment evaluates only the audio re-injection adapter. It does not
validate semantic-span commitment, irreversible tool-call gating, or
cross-provider generalization. The lookback ablation additionally compares
realtime batches, so its near-tie between 2.5 and 5~s warrants replication.
Human validation of RAA judgments, boundary-aware context selection, additional
runtime adapters, and natural conversational data are therefore the next steps.

%% file: tex/related_work.tex
\section{Related Work}\label{sec:related}

\noindent\textbf{Full-duplex spoken dialogue models.}\quad
Recent work distinguishes \emph{learned synchronization}, implemented inside a
spoken language model, from \emph{engineered synchronization}, implemented by
external dialogue components~\cite{fdslmsurvey2025}. Learned approaches remove
explicit turn boundaries in different ways. Moshi uses parallel speech
streams~\cite{moshi2024}; LSLM fuses dedicated listening and
speaking channels~\cite{lslm2024}; Synchronous LLMs incorporate wall-clock time
into generation~\cite{syncllm2024}; and NTPP jointly predicts token pairs from
dual-channel dialogue~\cite{ntpp2025}. Freeze-Omni and MiniCPM-o~4.5 further
combine speech understanding and generation with duplex training or unified
streaming~\cite{freezeomni2024,minicpm2026}. These models establish how an agent
can listen and speak concurrently. PACE addresses an orthogonal runtime issue:
how dialogue state should reflect the portion of a generated stream that has
actually reached playback.

\noindent\textbf{Controllers for cascaded full-duplex dialogue.}\quad
Engineered synchronization adds duplex behavior around cascaded or
semi-cascaded systems. Duplex Conversation uses user-state and barge-in
detection, while an LLM-based full-duplex scheme combines control tokens with a
neural finite-state machine~\cite{duplexconversation2022,fullduplexscheme2024}.
FlexDuo isolates duplex control in a plug-and-play module with an explicit idle
state, and LLM-enhanced dialogue management uses semantic control signals to
distinguish intentional from incidental barge-ins~\cite{flexduo2025,llmdm2025}.
More recent systems use micro-turns, personalized VAD, semantic end-of-turn
detection, and unit-level state transitions to retain the strengths of
ASR--LLM--TTS pipelines while supporting overlapping speech
\cite{duplexcascade2026,fireredchat2025,unitagent2026}. Their central task is
interaction arbitration: deciding whether and when the agent should yield,
resume, or respond. PACE consumes such a decision rather than replacing it; its
task begins afterward, by determining which assistant output remains valid for
the next model state or external action.

\noindent\textbf{Full-duplex dialogue evaluation.}\quad
Full-Duplex-Bench evaluates pause handling, backchanneling, turn-taking, and
interruption management~\cite{fullduplexbench2025}; FD-Bench extends the scope
to delay and robustness~\cite{fdbench2025}; subsequent versions and
MTR-DuplexBench add multi-turn tasks, instruction following, safety,
disfluencies, and tool use
\cite{fullduplexbenchv22025,mtrduplexbench2025,fullduplexbenchv32026}.
Other benchmarks target general spoken-dialogue understanding, task-oriented
interaction, and multi-aspect dialogue quality
\cite{urobench2025,tauvoice2026,mtalkbench2025}. These efforts primarily
measure interaction timing, task completion, or overall dialogue quality. To
cover the missing playback-relative dimension, we construct GCM-Bench. Its 108
cases vary the generation--playback gap and test whether referential requests
are anchored to played rather than merely generated output.

\noindent\textbf{Realtime interruption mechanisms.}\quad
Commercial realtime APIs and voice-agent frameworks expose the closest
deployed controls. OpenAI's Realtime API accepts an assistant item identifier
and an \texttt{audio\_end\_ms} offset in the
\texttt{conversation.item.}\allowbreak\texttt{truncate} event. It truncates the audio and removes the
corresponding server-side transcript so that unheard text does not remain in
the conversation~\cite{openaiRealtime}. Gemini Live cancels ongoing generation
when VAD detects an interruption, retains content already sent to the client,
and requires the application to stop playback and clear queued audio
\cite{geminiLiveDocs}. LiveKit combines endpointing, VAD, semantic or acoustic
turn detection, and adaptive interruption handling to distinguish intentional
barge-ins from backchannels and false interruptions~\cite{livekitTurns}.

These mechanisms demonstrate the practical importance of interruption-aware
state management, but expose different boundaries: item-relative audio time,
content sent to a client, or an interruption decision. We analyze the inherent
limitations of these item-level and state-level approaches in
Section~\ref{sec:state-surgery}. PACE factors out the common missing
abstraction---a turn-local playback boundary---and projects it through
whichever state-control interface the runtime provides.

\noindent\textbf{Conversational grounding.}\quad
Clark and Brennan describe grounding as the collaborative accumulation of
evidence that an utterance has been understood well enough for the current
purpose~\cite{clark1991grounding}; Clark subsequently frames language use as a
joint action coordinated by speakers and listeners~\cite{clark1996usinglanguage}.
PACE is not a cognitive model of attention or understanding; it uses playback
as a system-observable proxy for what the user heard. This evidence is
asymmetric: content that never reached playback could not have been heard,
whereas rendered content was available to be heard but was not necessarily
attended to or understood. PACE uses this conservative boundary to exclude
assistant content that cannot yet be treated as established dialogue context.

%% file: tex/discussion.tex
\section{Discussion}\label{sec:discussion}

\noindent\textbf{Scope of the current comparison.}\quad
The cancellation-only condition is intentionally a mechanism-level control: it
preserves the same model, interruption timing, and input audio while removing
the playback-context repair pipeline. The comparison therefore evaluates the
combined effect of re-injecting the played prefix and its spoken delimiter; it
does not separately identify their individual contributions. The delimiter
ablation confirms that this design choice materially affects anchoring. The
comparison should not be interpreted as establishing superiority over all
context-repair mechanisms. Stronger
alternatives include provider-native item truncation, aligned transcript
re-injection, and checkpoint-based state rewinding. These mechanisms require
runtime interfaces or aligned intermediate representations that are not
exposed by the black-box speech backend used in our prototype. To our
knowledge, no commercial black-box realtime speech backend currently exposes
these alternative context-repair mechanisms through interfaces that permit a
controlled, like-for-like comparison. We therefore leave such a comparison to
future work as suitable provider interfaces become available.

\noindent\textbf{Audio re-injection for native speech systems.}\quad
For a cascaded ASR--LLM--TTS pipeline, the played prefix can be transcribed and
reintroduced into the text history. This option is not universally available
in native speech-to-speech systems. Such systems may expose speech as the only
input modality, provide no external transcript, or offer no defined mechanism
for inserting text into an ongoing audio context. Relying on text re-injection
would therefore require additional assumptions about ASR availability,
audio--text alignment, and mixed-modal context handling. PACE instead repairs
the context through the modality that every system in its target class already
accepts: speech. Audio re-injection is thus a modality-preserving realization of
playback grounding under a weaker backend interface, rather than a claim that
audio must be intrinsically superior to text. For backends that support reliable text insertion, transcript re-injection remains a complementary option.

\noindent\textbf{Ambiguity of the \emph{next} operation.}\quad
The \emph{next} operation requires the playback anchor just as \emph{elaborate} and
\emph{repeat} do: a correct response must advance from the item at the playback
boundary. Where the three operations differ is in how observable this
requirement is. For \emph{elaborate} and \emph{repeat}, the played item alone
determines whether a response uses the correct referent. For \emph{next},
correctness additionally depends on the generated-but-unplayed continuation: a
response that advances from an unheard internal item still names a plausible
new item and is superficially indistinguishable from a correctly anchored one.
An LLM judge that observes only the played prefix and the post-interruption
response may therefore accept any novel, non-repeated item as ``next,'' even
when the underlying anchor is wrong. Reliable resolution of such cases may
require case-level human inspection against both the playback boundary and the
generated-but-unplayed sequence. We do not introduce this manual annotation
protocol in the current work; instead, we report \emph{next} separately, treat
it as weaker evidence, and leave human-validated evaluation of this operation
to future work.

%% file: tex/conclusion.tex
\section{Conclusion}\label{sec:conclusion}

LLM-based full-duplex voice dialogue can cause \emph{Generative Context
Mis-anchoring} (GCM) when model-facing context advances beyond what the client
has played. PACE addresses this mismatch with a provider-independent
middleware abstraction. It uses a turn-local playback boundary as an
observable proxy for what the user heard, keeps unheard output revocable, and
projects played context before post-interruption input. We implement its
physical, audio-only path in a browser-based assistant connected to a black-box
speech model without modifying the model service.

We also introduce GCM-Bench, a new 108-case dataset that we construct for
controlled playback-relative referent anchoring and publicly release to
support reproducible study of GCM. On GCM-Bench, PACE raises Referent
Anchoring Accuracy from 25.0\% to 96.3\%. In a separate study on 200 official FDB
v1 samples, the mean response score changes from 4.975 to 4.995, with 58.7~ms
higher mean latency only on the interruption path; requests made while the
assistant is not playing retain the original processing path and incur no
additional PACE latency. These
results demonstrate the feasibility of playback-grounded
context repair in practice.